\documentclass[a4paper,
               biblatex,     
               ]{jacow}
\usepackage{pdfpages,multirow,ragged2e} %
\usepackage{transparent}
\makeatletter%
	\ifboolexpr{bool{xetex}}
	 {\renewcommand{\Gin@extensions}{.pdf,%
	                    .png,.jpg,.bmp,.pict,.tif,.psd,.mac,.sga,.tga,.gif,%
	                    .eps,.ps,%
	                    }}{}
\makeatother

\ifboolexpr{bool{xetex} or bool{luatex}} 
 {}                                      

\usepackage[USenglish]{babel}

\usepackage{subfigure}
\usepackage{graphicx}
\usepackage{siunitx}
\usepackage{wasysym}

\usepackage{amsmath}
\usepackage{notoccite}

\usepackage{tabularx}
\usepackage{xspace}
\usepackage{float} 
\usepackage{placeins} 
\usepackage{upgreek}

\begin{document}

\renewcommand{\deg}{\textdegree C\xspace}
\newcommand{\midT}{\mbox{mid-T}\xspace}
\newcommand{\Q}{$Q_0$\xspace}
\newcommand{\Eacc}{$E_{acc}$\xspace}
\newcommand{\aqs}{anti-Q-slope\xspace}
\newcommand{\lowT}{low-T\xspace}
\newcommand{\EP}{electro polishing\xspace}
\newcommand{\hfqs}{high-field-Q-slope\xspace}
\newcommand{\bcurve}{black\xspace}
\newcommand{\mcurve}{blue\xspace}
\newcommand{\mlcurve}{purple\xspace}
\newcommand{\qe}{$Q_0(E_{acc})$\xspace}
\newcommand{\fig}{figure\xspace}
\newcommand{\figs}{figures\xspace}
\newcommand{\Ke}{K\xspace}
\newcommand{\Tab}{table\xspace}
\newcommand{\ol}{\textit{l}\xspace}
\newcommand{\rbcs}{$R_{BCS}$\xspace}
\newcommand{\rs}{$R_S$\xspace}


\title{Further improvement of medium temperature heat treated SRF cavities for high gradients\thanks{This work was funded by the Helmholtz Association within the MT ARD and the European XFEL R\&D Program.}}

\author{L. Steder\thanks{lea.steder@desy.de}, C. Bate, K. Kasprzak, D. Reschke,\\
 L. Trelle, H. Weise, M. Wiencek \\
Deutsches Elektronen-Synchrotron DESY, Germany \\
}
	
\maketitle
\begin{abstract}
The application of heat treatments on 1.3 GHz TESLA type cavities in ultra-high vacuum at 250\deg to 350\deg is called medium temperature or \midT heat treatment.
In various laboratories such treatments on  superconducting radio frequency (SRF) cavities result reproducible in three main characteristic features for the quality factor \Q in dependency of the accelerating electric field strength~\Eacc.
First, comparing \midT heat treatment with a baseline treatment,
a significant increase of \Q up to $5\cdot10^{10}$ at 2 \Ke can be observed.
Second, with increasing accelerating gradient \Eacc the \Q increases up to a maximum around 16 to 20 MV/m. This effect is known as \aqs. The third observation for a \midT heat treatment compared to a baseline treatment is an often  reduced maximum gradient \Eacc.

In \cite{DESYmidT} the appearance of a \hfqs (HFQS) is reported after \midT heat treatments of 3 hours at 350\deg or of 20 hours at 300\deg at DESY.
Using the heating temperature and the heating time taken from the temperature profile of the furnace effective oxygen diffusion lengths \ol were calculated.
In the follow-up study presented here, a set of three single-cell cavities with diffusion lengths \ol above 1700 nm, showing HFQS, were treated with an additional so-called \lowT bake of 24-48 hours at 120\deg to 130\deg. 
The subsequent reproducible \qe -performances results indicate that the \lowT bake procedure cures the HFQS like for cavities treated with the EuXFEL recipe \cite{ReschkeXFEL} of \EP (EP) and following \lowT treatments.
As presented in the following, \Q values of more than $3\cdot10^{10}$ at 16 MV/m and accelerating gradients of 32 to 40~MV/m are achieved.
More detailed analyses of the cavity performances - especially their sensitivity against trapped magnetic flux - as well as the application to EuXFEL type nine-cell cavities are currently under preparation.
\end{abstract}

\section{Combination of heat treatments for high gradients}
For a future upgrade of the European XFEL an exchange of the first 17 accelerator modules is proposed \cite{JacekCW}. The performance goals for the new cavities are envisaged with a \Q of $2.7\cdot10^{10}$ at \Eacc of 16~MV/m. 
Nowadays, the high-duty-cycle (HDC) working group which is coordinating the R\&D work towards the mentioned upgrade, even proposes a \Q above $3\cdot10^{10}$ and gradients larger than 20~MV/m \cite{HDC} for the HDC operation mode. In addition, the EuXFEL shall be still be operable in the pulsed mode for high beam energies, for which large accelerating gradients of the cavities are needed as well. 
The present cavity R\&D activities at DESY are therefore focused on \midT heat treatments.

In \cite{DESYmidT}, the results of 19 heat treatments on single-cell 1.3~GHz TESLA cavities, which were treated in UHV (ultra-high vacuum) at medium temperatures of 250\deg to 350\deg with a duration between 3 and 20 hours, were analyzed.
Five of the used single-cell cavities are fabricated of large-grain (LG) niobium material, all others are fine-grain (FG) cavities.
In the following, the most important findings of \cite{DESYmidT} are shortly reported and the scope of the additional analysis presented in this very paper are introduced.

\begin{figure}[H]
   \centering
   \includegraphics*[width=1.0\columnwidth]{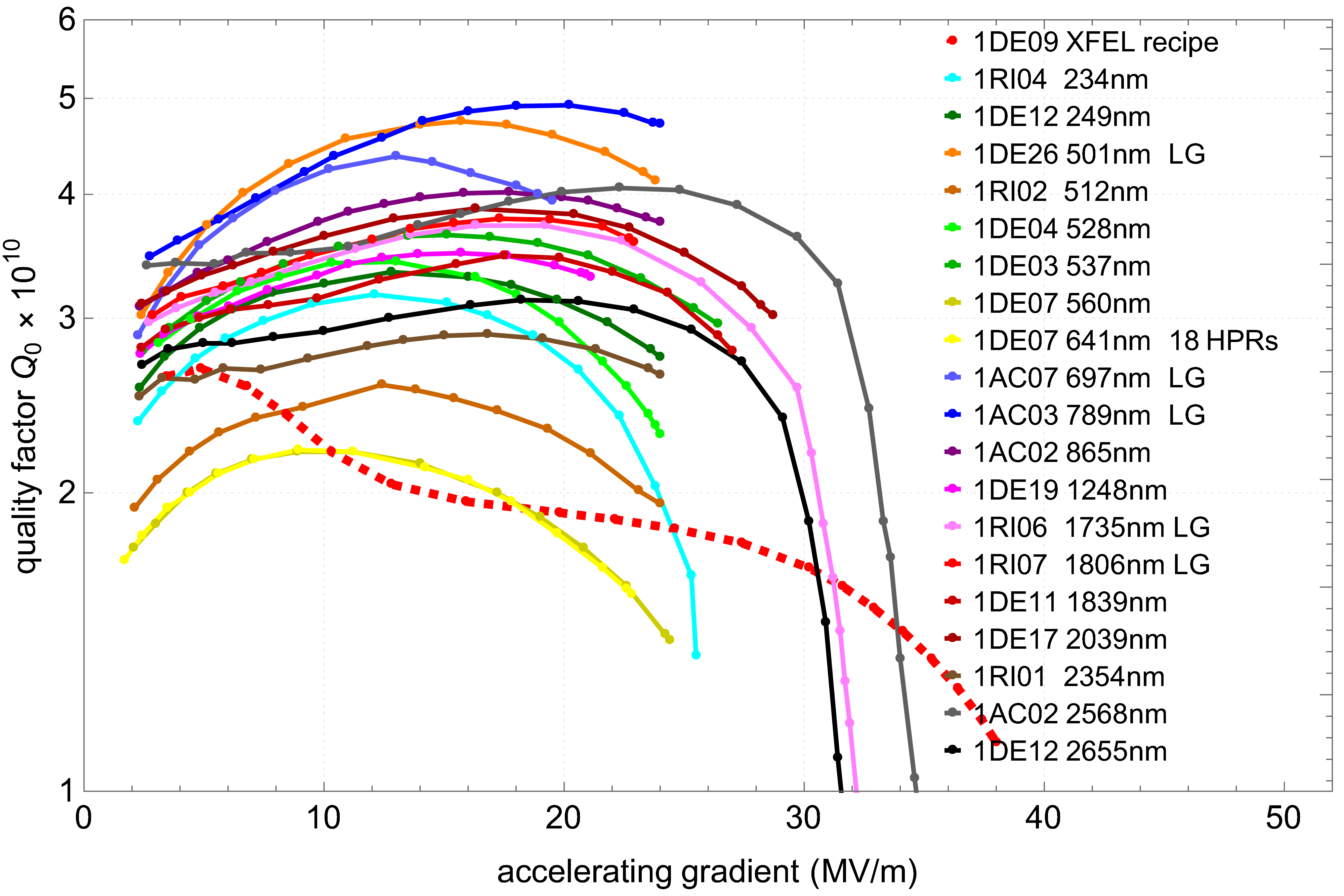}
   \caption{\qe curves of all \midT heat treated cavities presented in \cite{DESYmidT} and comparison to a single-cell cavity prepared with the standard EuXFEL recipe \cite{ReschkeXFEL} consisting of \EP followed by \lowT bake. The treatments are labeled by their oxygen diffusion lengths, the treatment parameters (temperature and duration) can be found in \cite{DESYmidT}.}
   \label{fig:allmidT}
\end{figure}

The \midT heat treatment requires no additional gases like nitrogen during the process, is highly reproducible and eliminates the need for a subsequent chemical surface treatment.
In the analysis the treatments were categorized according to the effective oxygen diffusion length \ol based on the whole temperature vs. time profile of the furnace treatments, more details are explained in \cite{DESYmidT}. 
As can be seen in \fig \ref{fig:allmidT}, a characteristic increase of the quality factor over \Eacc with a \Q maximum at 16 to 20 MV/m can be observed for all treatments.
A significant reduction of the BCS surface resistance \rbcs was observed as well.
For the investigated range of the oxygen diffusion length \ol between 234~nm and 2655~nm, the \Q(\Eacc=16 MV/m) at 2\Ke was independent of \ol. 

\begin{figure}[htb]
   \centering
   \includegraphics*[width=1.0\columnwidth]{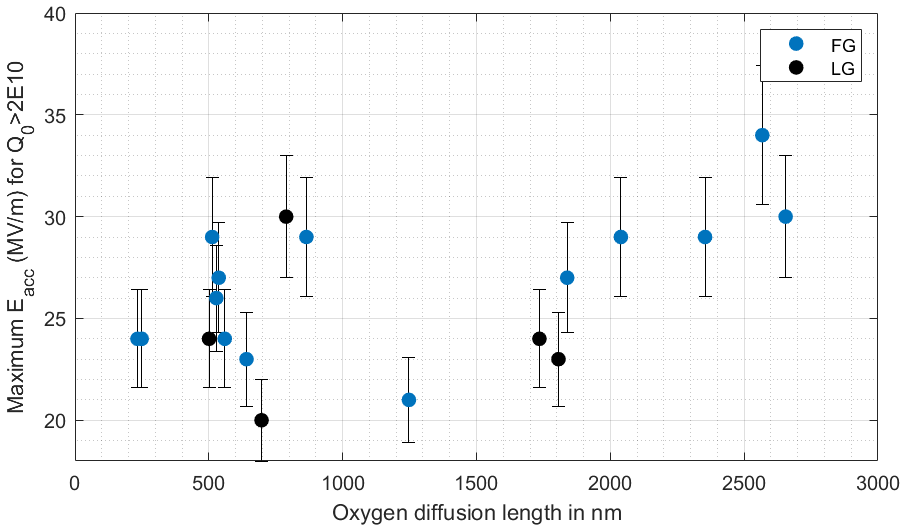}
   \caption{Maximal achieved gradient \Eacc for
all mid-T heat treatments with \Q $>2\cdot10^{10}$ versus oxygen diffusion length \ol, \cite{DESYmidT}.}
   \label{fig:E_l}
\end{figure}

For most treatments, no gradient \Eacc higher than 30~MV/m could be achieved, as can be seen in \fig \ref{fig:E_l}. This observation is consistent with the results of other laboratories  \cite{ItoKEK,HeIHEP,SHINE}.
Interestingly, a slight trend towards higher gradients with larger diffusion length \ol was found as can be seen in the right part of \fig \ref{fig:E_l}.

Moreover, for large oxygen diffusion lengths \ol~-- corresponding to temperatures around 300\deg for 20h and around 350\deg for 3h -- the so-called \hfqs (HFQS) with its characteristic exponential decay of the \Q -value above an onset accelerating field of about 28 MV/m was observed.  
The HFQS is a well-known feature of the \qe -performance after an \EP as final surface treatment \cite{GigiSRF} or a so-called 'soft reset' via 800\deg annealing \cite{DESYmidT}.
Typically such curves showing HFQS are mostly limited by available RF power (PWR) between 30 and 35 MV/m, meaning the cavity can not be driven to the breakdown (BD) of the superconductivity, which is also called quench.

Nevertheless, the HFQS was not expected to occur after a heat treatment at such low temperatures around 350\deg.
The well-established empirical procedure to cure the HFQS is a so-called '\lowT bake'. 
Several causes for the \hfqs and methods to overcome it via \lowT heat treatments are reviewed in \cite{Ciovati120}.
An additional explanation for the occurrence of HFQS is possible via so-called 'Nano-hydrids', details can be found in \cite{NanoFNAL}.

During such a \lowT bake procedure an UHV inside of the cavity is maintained and a wide range of parameters for temperature (90\deg - 150\deg) and duration (12h - 100h) can be applied.
Typically a 48 hours at 120\deg process is used, as in the cavity series production for the EuXFEL \cite{ReschkeXFEL}.
In 2018 it was observed that the so-called 'two-step-bake' adding four hours at 75\deg at the beginning of a standard \lowT procedure of 48 hours at 120\deg \cite{twoStepFNAL} enhances the cavity gradient \Eacc.
A statistical analysis of EuXFEL cavity data \cite{StederLowT} and the above mentioned two-step-bake led to the implementation of a process of 4 hours at 75\deg followed by 24 hours at 130\deg as standard \lowT treatment at DESY.

In order to investigate the response of the three \midT heat treated single-cells showing \hfqs, an additional \lowT bake process and a subsequent vertical test were performed.

\section{Experimental overview}
The DESY furnace infrastructure, used for \midT treatments, consisting of a refurbished niobium retort furnace located in the cavity assembly cleanroom as well as its control features and the related workflow are described in detail in \cite{DESYmidT,TrelleFurnace}.

The complete information about the used single-cell cavities, their surface treatments and niobium material, the testing environment and procedures as well as further explanations of the measured variables can be found in \cite{DESYmidT}.

The measurement uncertainty for independent RF measurements is approximately 10\% for \Eacc and up to 20\% for \Q. 
However, within a single vertical test and for each curve, the observed measurement deviation is significantly smaller, only around 1\% for \Eacc and 3\% for \Q \cite{ReschkeXFEL}.

Results of three cavities are presented in the following, they are called 1AC02, 1DE12 and 1RI06 - with the latter made of LG niobium. All of them received a baseline treatment in form of a soft reset via 800\deg annealing or a short \EP and a consecutive vertical performance test. 
In between each baseline or \midT heat treatment and these vertical tests a standard cycle of six high-pressure rinsings (HPRs) followed by the assembly of auxiliaries and especially an adjustable antenna for adaptive coupling takes place. 
Furthermore \midT heat treatments resulting in large oxygen diffusion lengths were applied according to the parameters given in column two of \Tab \ref{tab:lowT} and followed by an extensive vertical test. 
\begin{table}[htb]
    \tabcolsep=0.11cm 
    \begin{tabular}{c|c|c|c}
    \toprule
    cavity & \midT & \lowT & condition  \\
    & process & process & @ \lowT \\
    \hline
    1AC02  & 3h@350\deg & 4h75\deg 24h120\deg & pumped\\
    1RI06 LG& 20h@300\deg & 4h75\deg 24h120\deg & pumped\\
    1DE12 & 3h@350\deg & 4h75\deg24h120\deg & open in Ar\\
    \bottomrule
   \end{tabular}
   \caption{Different cavity \midT and \lowT heat treatments.}
    \label{tab:lowT}
\end{table}

Due to the appearance of the \hfqs for these three cavities, two of them received the \lowT bake in the standard environment in an inert gas atmosphere on the fully assembled and evacuated cavity ready for vertical test. 
1DE12 was treated 'open' without any attached flanges under argon atmosphere in a dedicated heating chamber
which is located in the DESY cleanroom. The design of this chamber is based on the successful initial experiments reported in \cite{ReschkeCabinet}.
Since this chamber is only usable up to a maximal temperature of 120\deg, also the other \lowT heat treatments were restricted to this maximum temperature.
Parameters of the different applied \lowT bake procedures can also be found in \Tab \ref{tab:lowT}.

\section{Results of \midT and \lowT heat treated SRF cavities}
In the scope of the recent \midT campaign at DESY seven single-cell cavities were treated for either 3 hours at 350\deg or 20 hours at 300\deg resulting in calculated oxygen diffusion lengths in the range of 1700~nm to 2700~nm \cite{DESYmidT}.
Three of them showed a clear \hfqs behaviour after the \midT heat treatment.
Hence, they are the subject of \lowT treatment studies in the following. 

\subsection{Quality factors and accelerating gradients}
The onset of the \hfqs of the aforementioned cavities ranges between 28~MV/m and 32~MV/m.
Above these field strengths \Q decreases exponentially with increasing accelerating gradient, which can be seen in the \mcurve curves in the following \figs \ref{fig:1AC02}, \ref{fig:1RI06} and \ref{fig:1DE12}. 
All of the here shown measurements, no matter whether before or after heat treatments,  were performed field emission free and at an operation temperature of 2\Ke.

In addition, also the baseline performances after the reset treatments are shown in \bcurve.
All of the baseline measurements exhibit the typical \hfqs after \EP or 800\deg treatments. The end of the \bcurve curves is always determined by reaching the RF power limit during vertical testing.

Comparing the performances before (\bcurve) and after (\mcurve) the \midT heat treatments, the typical features like significant \Q enhancement and \aqs can be observed for all three cavities.
The \midT heat treatment characteristic of an early quench is not occurring, instead a \hfqs can be observed.
In contrast to the baseline measurements, the measurements of the \midT heat treatments are limited by quenches and not by an RF power limit.

The most interesting curves in \figs \ref{fig:1AC02}, \ref{fig:1RI06} and \ref{fig:1DE12} are the \mlcurve ones, since they show the cavity performances after the additional \lowT bake procedures.
For all three cavities, the \hfqs after the \midT heat treatment is cured by the additionally applied \lowT treatments.
All of the \mlcurve curves are limited by quenches of the cavities, which occur between 32 MV/m for 1RI06 and 40 MV/m for 1AC02.

\subsubsection{1AC02}
\begin{figure}[htb]
   \centering
   \includegraphics*[width=1.0\columnwidth]{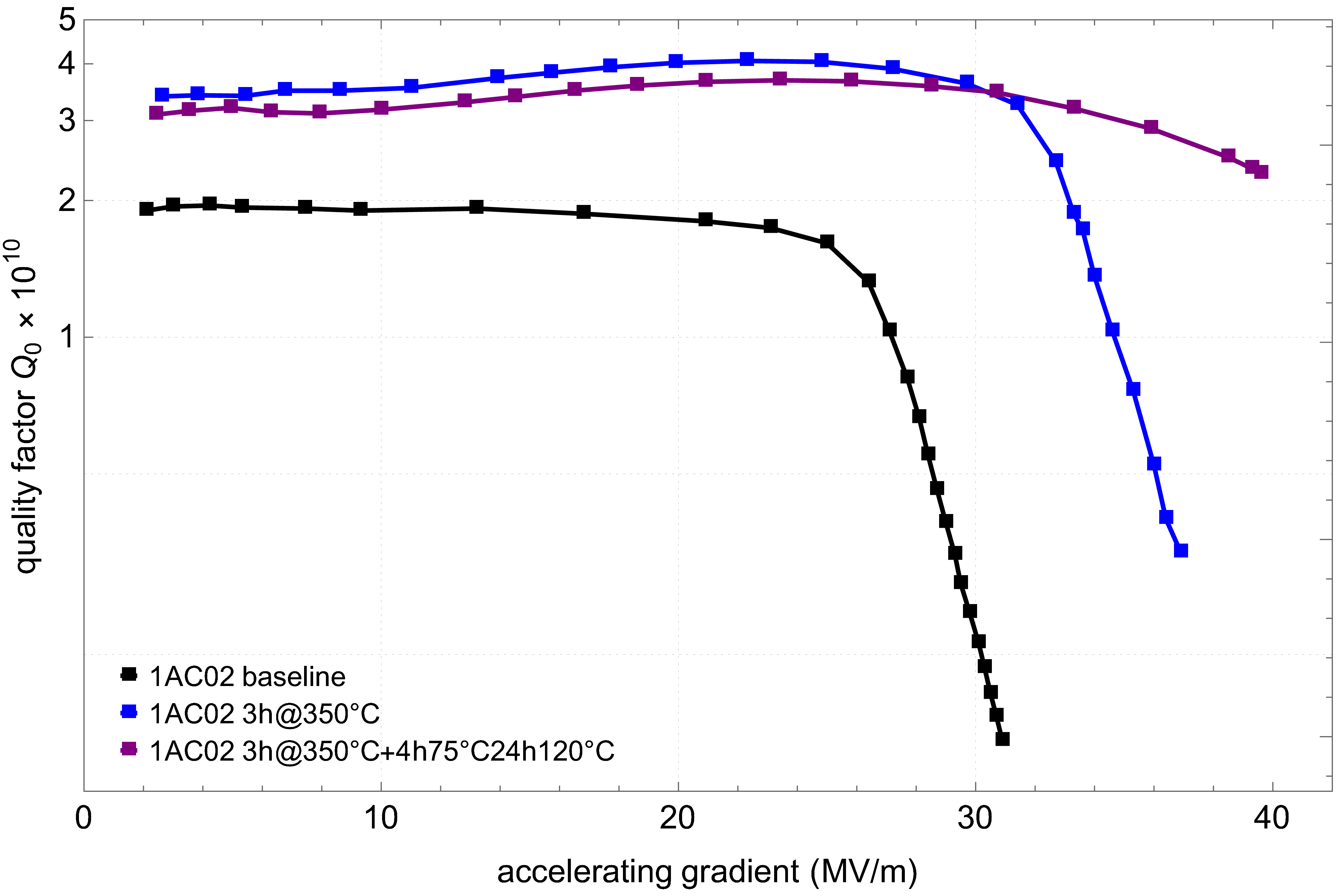}
   \caption{\qe curves of cavity 1AC02 at 2\Ke. In \bcurve the performance after the baseline treatment (3h 800\deg) is shown (limited by PWR), while the \mcurve data depicts the behaviour after the \midT heat treatment of 3h@350\deg (limited by BD). The \mlcurve curve shows the condition after the final \lowT baking for 4h75\deg24h120\deg under vacuum (limited by BD).}
   \label{fig:1AC02}
\end{figure}

In \fig \ref{fig:1AC02} the \qe behaviour of 1AC02 after the baseline annealing (\bcurve), the \midT heat treatment of 3h@350\deg in the niobium retort furnace (\mcurve) and the additional \lowT procedure (\mlcurve), which took place in a UHV pumped condition, is shown.
During both tests after the \midT and \lowT treatment a possible antenna overheating was observed for the last few measurement points, while applying high RF power.
A re-test with exchanged adjustable antenna is under preparation.

The very large gain in \Q produced by the \midT heat treatment is almost preserved after the \lowT bake.
Most prominent is the very large maximal accelerating gradient \Eacc of 40 MV/m with a \Q of $2.4\cdot10^{10}$ at this point. 

\subsubsection{1RI06}
\begin{figure}[htb]
   \centering
   \includegraphics*[width=1.0\columnwidth]{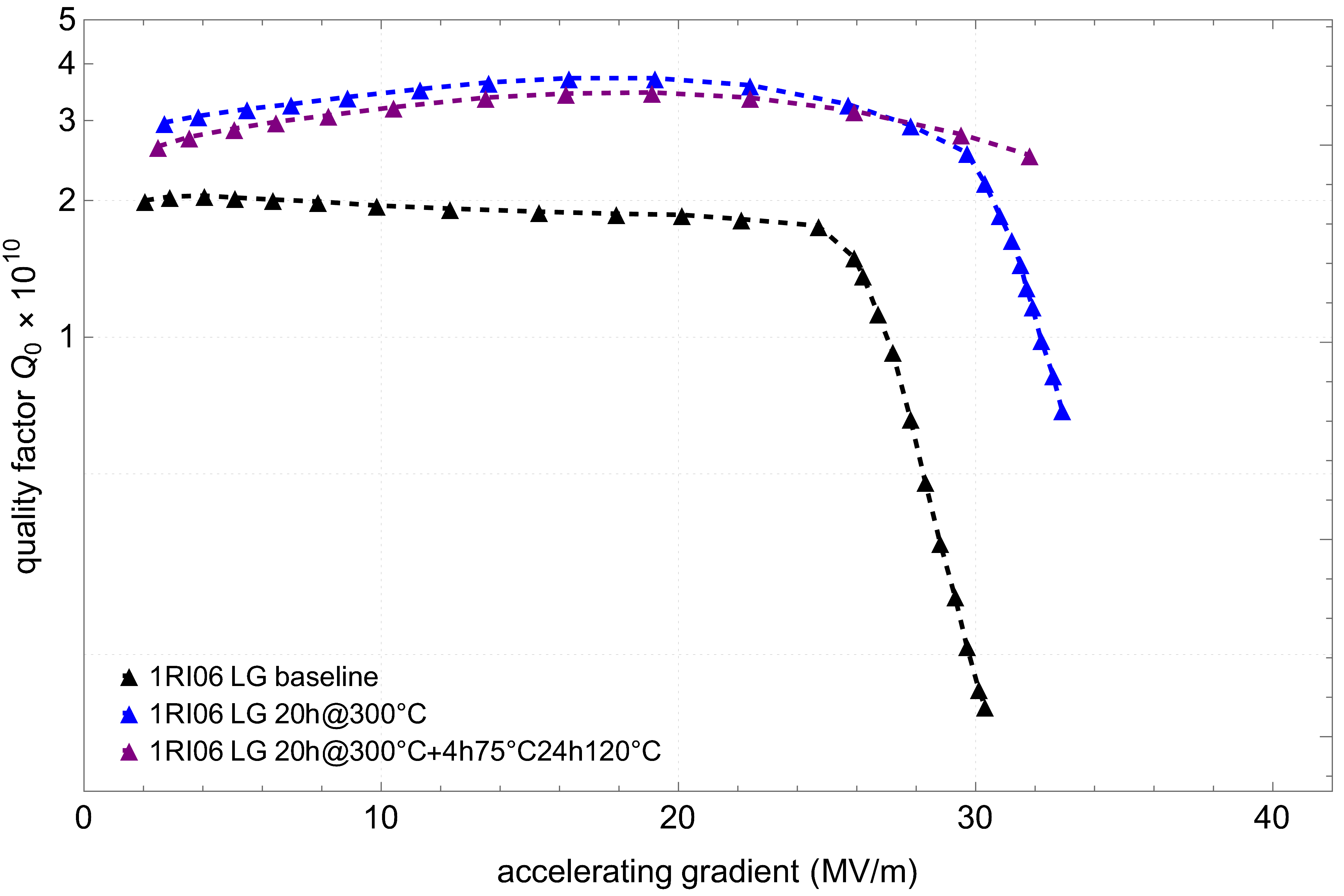}
   \caption{\qe curves of large grain cavity 1RI06 at 2\Ke. In \bcurve the performance after the baseline treatment (20 $\upmu$m EP) is shown (limited by PWR), while the \mcurve data depicts the behaviour after the \midT heat treatment of 20h@300\deg. The \mlcurve curve shows the condition after the final \lowT baking for 4h75\deg24h120\deg under vacuum (limited by BD).}
   \label{fig:1RI06}
\end{figure}

The performances of the large grain cavity 1RI06 are shown in \fig \ref{fig:1RI06}. After its production the cavity got a final EP with a material removal of 20 $\upmu$m  before the \midT heat treatment procedure of 20h@300\deg. This prolonged treatment at 300\deg results in similar oxygen diffusion lengths as 3h 350\deg procedures \cite{DESYmidT}.  

\begin{figure*}[hbt]
    \centering
        \subfigure[]{\includegraphics*[width=1\columnwidth]{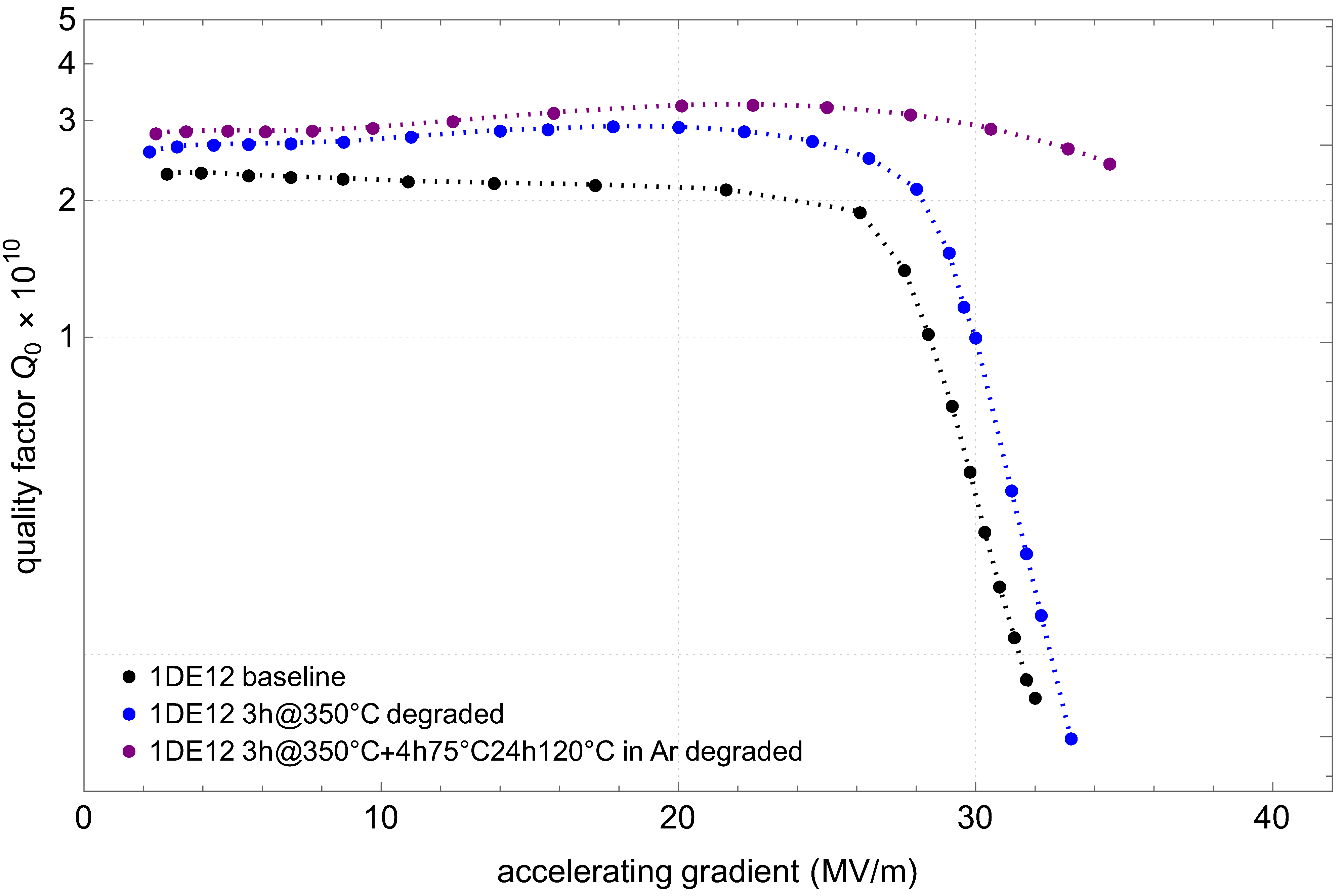}}
        \subfigure[]{\includegraphics*[width=1\columnwidth]{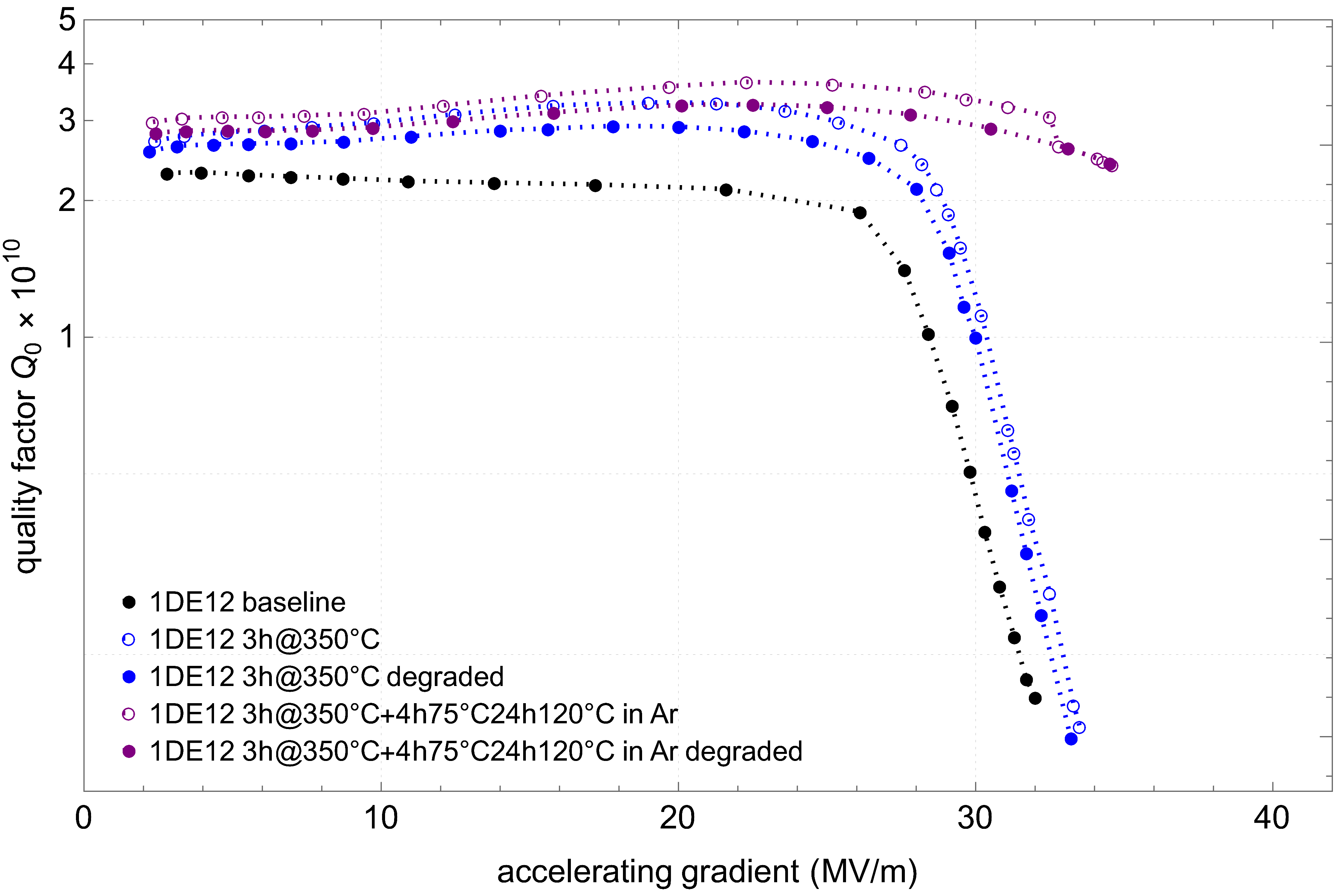}}
    \caption{\qe curves of cavity 1DE12 at 2\Ke. In \bcurve the performances after the baseline treatment (3h 800\deg) are shown (limited by PWR), while the \mcurve data depict the behaviour after the \midT heat treatment of 3h@350\deg in the niobium retort furnace (limited by BD). The \mlcurve curves show the condition after the final open \lowT baking for 4h75\deg24h120\deg in argon atmosphere (limited by BD). In figure (a) the degraded performances after \midT and \lowT treatments are shown here, while in (b) the not-degraded performances are visible as well.}
     \label{fig:1DE12}
\end{figure*}
%
%
Also here, the enormous difference between the baseline curve in \bcurve and the performance after the heat treatments is obvious and demonstrates the advantage of \midT heat treatments.
By curing the \hfqs with a \lowT bake under vacuum, the accelerating gradient is in this case not enhanced.
But with a \Q of $2.4\cdot10^{10}$ at 32 MV/m also here an outstanding performance can be observed.
This cavity degrades in the quality factor after first quenching. 
As described in \cite{DESYmidT}, this degradation can be healed via a thermal cycling up to 30\Ke.
After reaching the operation temperature of 2\Ke again, the initial higher \Q value is restored.

\subsubsection{1DE12}
In contrast to the other two single-cells, cavity 1DE12 received its \lowT heat treatment in a dedicated heating chamber in the DESY cleanroom while being exposed to argon atmosphere.
Also for this cavity a degradation in \Q can be observed. In this case, the degraded performances are shown in \fig \ref{fig:1DE12} (a), since they are better comparable.
Additionally, the non-degraded \qe curves are shown in \fig \ref{fig:1DE12} (b) to enable an evaluation of the possible \Q factor.
The curve of the first measurement (\mlcurve, open markers) shows the degradation in \Q (kink) before the maximal achievable gradient. 
This was not only observed here, but also in other vertical tests of \midT heat treated cavities, which were performed at DESY.
A further analysis, which may reveal an explanation for the degradation process needs to be conducted in future.
A quality factor of $2.4\cdot10^{10}$ at 35 MV/m even in the degraded state is again a very good performance of a cavity after the combination of \midT and \lowT heat treatment.
Furthermore, it is notable that the quality factor is even slightly improved after the \lowT baking process compared to the \midT heat treatment

\subsubsection{Performance comparison}
\begin{figure}[H]
   \centering
   \includegraphics*[width=1.0\columnwidth]{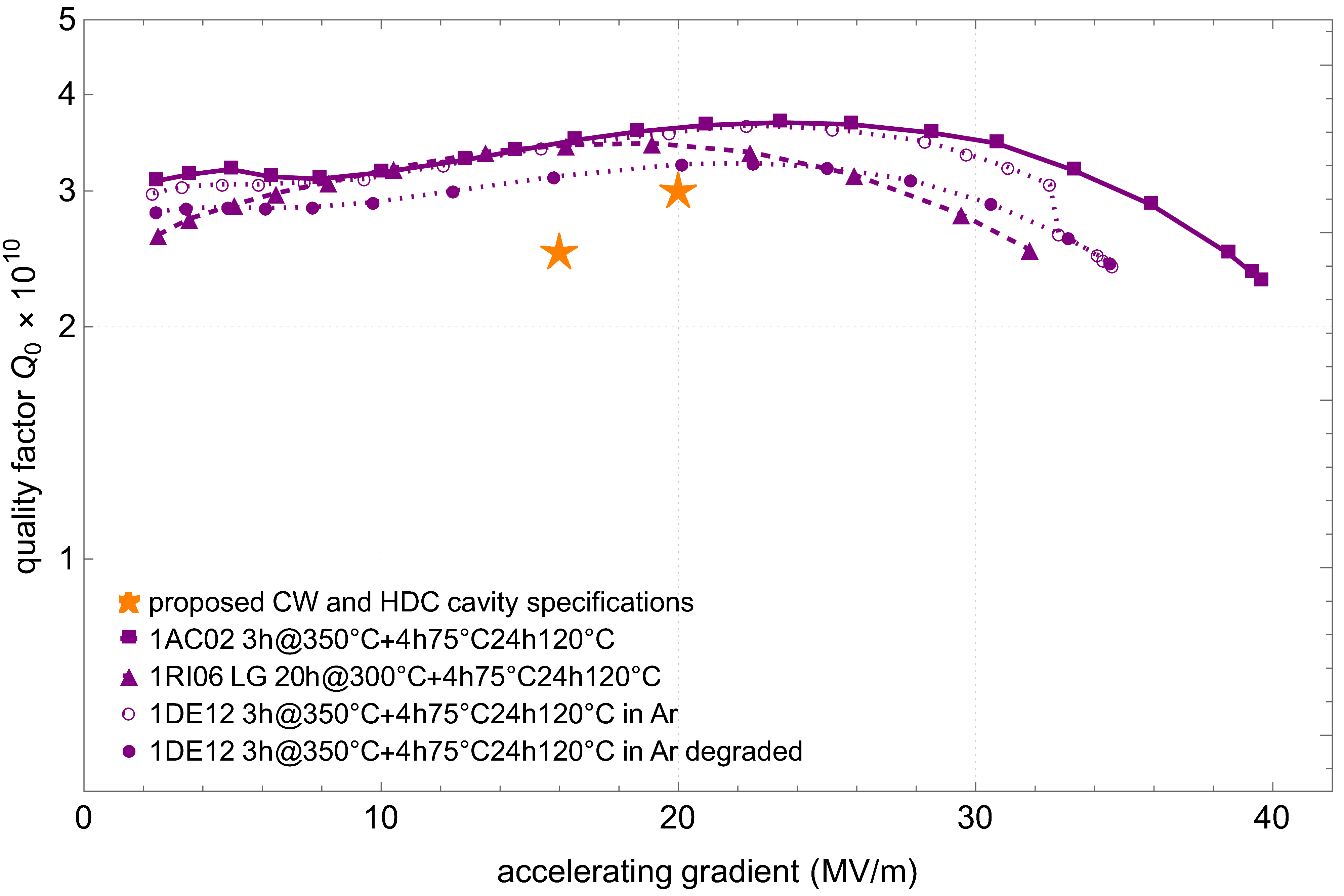}
   \caption{\qe curves for three single-cell cavities at 2\Ke after the \midT --\lowT heat treatment combination. The stars indicate the two suggested specification points for EuXFEL upgrade cavities.}
   \label{fig:QE_3}
\end{figure}

In \fig \ref{fig:QE_3} the \qe curves of all three cavities after \midT --\lowT treatment combination are compared. 
It is obvious that the performances of all three of them are outstanding and very promising since large accelerating gradients are in reach.
All quality factors are above $2.4\cdot10^{10}$ over the complete \Eacc range at 2\Ke. At 16~ MV/m and 20~MV/m respectively the values are between $3.2\cdot10^{10}$ and  $4.0\cdot10^{10}$ thus significantly well above the foreseen specification for an European XFEL upgrade.

\subsection{Surface resistances}
In order to learn more about the underlying processes which lead to the improved performance after \lowT bake appliance, the composition of the surface resistance $R_{S}(T,B)$ is analyzed.
The surface resistance of a SRF cavity can be written as 
\begin{equation}
    R_S(T, B) = R_{BCS}(T) + R_{const}, 
\end{equation}
where $R_{BCS}(T)$ depicts the temperature dependent BCS resistance and $R_{const}=R_{res}+R_{flux}(B)$ consisting of the temperature independent parts: The residual resistance $R_{res}$ and $R_{flux}(B)$, an additional surface resistance induced by magnetic flux.
Both \rbcs and $R_{const}$ are studied separately in the following.

An estimation of \rbcs at 2\Ke can be gained using \qe curves taken at 2\Ke and 1.5\Ke with $Q_0 \approx 1/R_S$ and $R_{BCS,2~\text{K}} \approx R_{S,2~\text{K}}-R_{S,1.5~\text{K}}$.

\begin{figure}[H]
   \centering
   \includegraphics*[width=1.0\columnwidth]{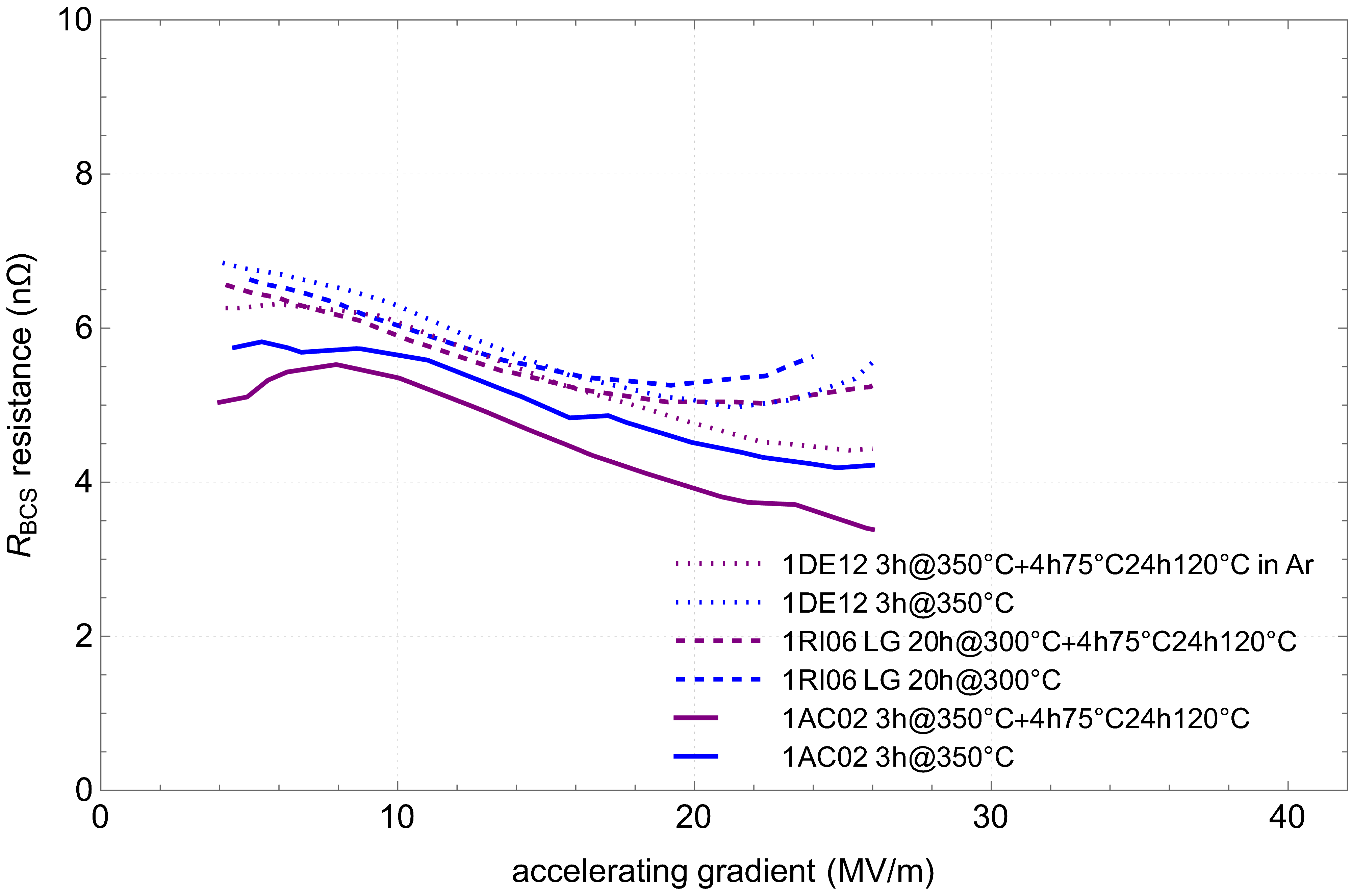}
   \caption{Estimation of $R_{BCS}$ with the help of $R_{BCS,2~\text{K}}~\approx~ R_{S,2~\text{K}}-R_{S,1.5~\text{K}}$ for all three \midT--\lowT heat treated cavities. Due to HFQS, the evaluation is stopped at 26 MV/m.}
   \label{fig:RBCS_3}
\end{figure}

The latter equation is used to determine \rbcs curves for all three cavities after \midT and after the combination of \midT and \lowT heat treatments which are shown in \fig \ref{fig:RBCS_3}.
Due to the steep \hfqs in the \mcurve curves, the evaluation beyond the value of 26~MV/m makes no sense.
Interestingly the BCS resistance is reduced in all cases after the additional applied \lowT procedure, though the significance is not good, due to uncertainties in the range of 1 to 2 n$\Omega$ \cite{DESYmidT}. 

The constant part of the surface resistance can be approximated by the evaluation of \qe curves at 1.5\Ke and in this case at 16 MV/m.
The corresponding results can be found in \Tab \ref{tab:Rconst}. While the temperature dependent resistance is reduced after the \lowT treatment (cf. \fig \ref{fig:RBCS_3}), the here shown temperature independent part appears to be increased.
Only for 1DE12 the $R_{const}$ is lowered, but this change of 0.3 n$\Omega$ is smaller than the uncertainty level. 
\begin{table}[H]
    \centering
    \tabcolsep=0.11cm 
    \begin{tabular}{c|c|c}
    \toprule
    cavity & $R_{const}$ & $R_{const}$\\
    & after \midT & after \lowT \\
    \hline
    1AC02  & 2.2 n$\Omega$ & 3.4 n$\Omega$\\
    1RI06 LG & 1.9 n$\Omega$ & 2.7 n$\Omega$\\
    1DE12 & 3.0 n$\Omega$ & 2.7 n$\Omega$\\
    \bottomrule
   \end{tabular}
   \caption{Temperature independent resistance contributions at accelerating gradients of 16 MV/m for cavities before and after additional \lowT bakes.}
    \label{tab:Rconst}
\end{table}

Here, only observations are reported, while for the interpretation of the results more studies with cavities subjected to the combination of \midT and \lowT heat treatments are necessary.

\section{Summary}
Applying a \midT heat treatment of either 3 hours at 350\deg or 20 hours at 300\deg results in the cavities showing characteristic \midT performances but also a \hfqs. This slope can be cured by an additionally applied \lowT procedure. 

The newly developed \midT --\lowT chain leads to quality factors well above $2\cdot10^{10}$ over the complete gradient range and in the region of interest for a possible EuXFEL upgrade even values between $3\cdot10^{10}$ and $4\cdot10^{10}$ are reached.

For the first time, \midT heat treated single-cell cavities achieve, after an additional \lowT bake, besides their large quality factors \Q also accelerating gradients \Eacc of up to 40~MV/m. 

The behaviour of the \rbcs and $R_{const}$ needs to be analyzed in more detail in order to optimize the process even further.

\section{Acknowledgements}
We express our gratitude to the DESY and University of Hamburg SRF team for fruitful discussions and especially the cleanroom staff and the AMTF team for their essential support during cavity preparation and testing. This work was supported by the Helmholtz Association within the topic Accelerator Research and Development (ARD) of the Matter and Technologies (MT) and benefits greatly from the European XFEL R\&D Program.

\printbibliography

\end{document}